# Switchable Organic Plasmonics with Conductive Polymer Nanoantennas

**Shangzhi Chen[1], Evan S. H. Kang[1], Mina S. Chaharsoughi[1], Vallery Stanishev[2], Philipp Kühne[2], Hengda Sun[1], Vanya Darakchieva[2] and Magnus P. Jonsson[1]★**

[1]Laboratory of Organic Electronics, Department of Science and Technology (ITN), Linköping University, SE-601 74 Norrköping, Sweden. [2]Terahertz Materials Analysis Center (THeMAC) and Center for III-N Technology, C3NiT - Janzèn, Department of Physics, Chemistry and Biology (IFM), Linköping University, SE-581 83 Linköping, Sweden.
★e-mail: magnus.jonsson@liu.se

**Metal nanostructures are key elements in nanooptics owing to their strong resonant interaction with light through local plasmonic charge oscillations. Their ability to shape light at the nanoscale have made them important across a multitude of areas, including biosensing[1], energy conversion[2] and ultrathin flat metaoptics[3]. Yet another dimension of avenues is foreseen for dynamic nanoantennas[4], ranging from tuneable metalenses for miniaturized medical devices to adaptable windows that control radiation flows in and out of buildings. However, enabling nano-optical antennas to be dynamically controllable remains highly challenging and particularly so for traditional metals with fixed permittivity[4]. Here we present state-of-the-art conductive polymers as a new class of organic plasmonic materials for redox-tuneable nano-optics. Through experiments and simulations, we show that nanodisks of highly conductive polymers can provide clear optical extinction peaks via excitation of dipolar localised surface plasmon resonances. Resonance frequencies redshift with increasing nanodisk aspect ratio, in agreement with analytical calculations based on dipolar polarizability theory. We furthermore demonstrate complete switching of the optical response of the organic nanoantennas by chemical tuning of the polymer's redox state, which effectively modulates the material permittivity between plasmonic and non-plasmonic regimes. Our results thereby show that conductive polymer nanostructures can act as redox-tuneable plasmonic nanoantennas, based on bipolaronic charge carriers rather than electrons as in conventional metals. Future directions may investigate different polymers and geometries to further widen the plasmonic spectral range (here around 0.8 to 3.6 µm) as well as different ways of tuning.**

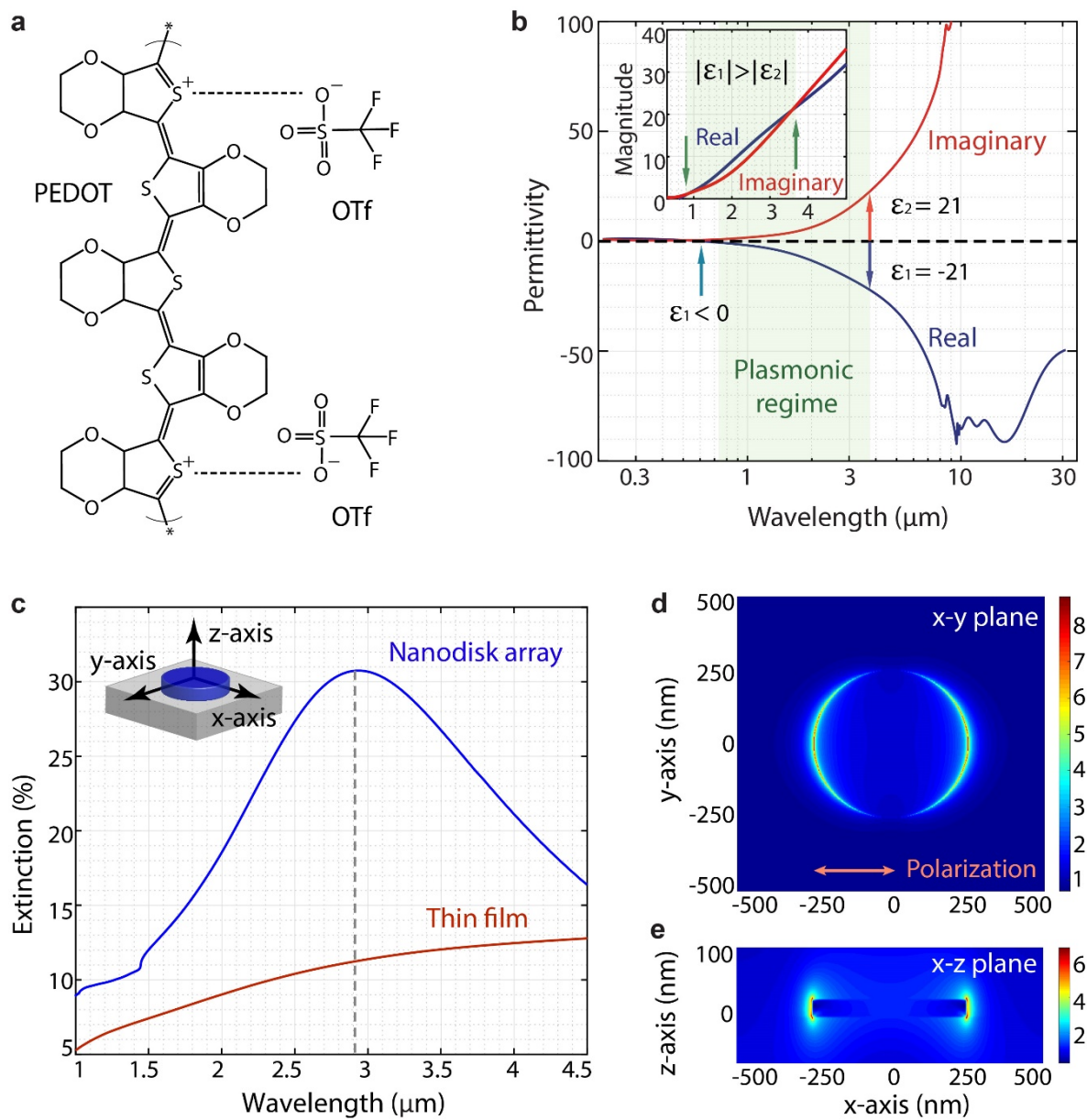

**Fig. 1 | Material properties and calculated plasmonic resonances for PEDOT:OTf in its high-conductivity oxidized state. a,** Chemical structure of PEDOT:OTf. **b,** In-plane permittivity dispersion of PEDOT:OTf in its oxidized state (blue curve: real part; red curve: imaginary part). The shaded spectral range between 0.8 to 3.6 µm is defined as plasmonic regime where the real permittivity is below zero and its magnitude is larger than the imaginary component (inset: absolute magnitude of real and imaginary permittivity). **c,** Simulated extinction spectrum for a PEDOT:OTf nanodisk array (blue curve), with disk thickness of 30 nm, diameter of 500 nm, and periodicity of 1000 nm. The red curve shows the extinction for a non-structured thin PEDOT:OTf film scaled to the same material coverage as the disks (scaled by π/16). Inset: a schematic illustration of a PEDOT:OTf nanodisk (blue) on glass substrate (grey) with indicated *x*-, *y*-, and *z*-axes. **d, e,** Calculated nearfield profiles at the wavelength of the extinction maximum (2.9 µm) for one of the PEDOT:OTf nanodisks of the array in **c**: **d**, *x-y* in-plane direction 2 nm above the nanodisk; **e**, *x-z* cross-section through the center of the nanodisk. The color scale bars show the electric field strength relative to the incident light.

Conductive polymers are conjugated materials with electrical conduction originating from (bi-) polaronic charge carriers situated along their backbones.[5] Materials based on poly[3,4-ethylenedioxythiophene] (PEDOT), especially PEDOT doped with trifluoromethanesulfonate (PEDOT:OTf), have shown particularly high electrical conductivity and metallic character[6,7]. We base our study on PEDOT:OTf (chemical formula in Fig. 1a) thin films produced by vapour phase polymerization (see Methods) and treated by sulphuric acid to further enhance the conductivity beyond 5000 S/cm (see Table. S1). The possibility for resonant light-matter interaction in these materials is governed by their complex permittivity $\varepsilon(\lambda)$, which we determined by ultrawide range spectroscopic ellipsometry, employing an anisotropic Drude-Lorentz model as described previously[8]. Fig. 1b shows the resulting in-plane permittivity of thin PEDOT:OTf film with thickness of 32 nm (Fig. S1 presents the raw data). The shaded area highlights a spectral region (0.8 to 3.6 µm) with negative permittivity $\varepsilon_1$ (blue curve) that is also larger in magnitude than that of the imaginary component $\varepsilon_2$ (red curve), which we refer to as plasmonic regime. The optically metallic and plasmonic character is related to the high conductivity within the thin film, in turn, due to high concentration ($2.6 \times 10^{21}$ cm$^{-3}$) of mobile bipolaronic charge carriers. We also note that the mobility is highly anisotropic[8,9] and the out-of-plane permittivity (Fig. S2a) is primarily positive throughout the measured range, making the conductive polymer thin film a natural hyperbolic material[10] (Fig. S3).

The measured optical properties of the thin PEDOT:OTf film imply that nanostructures of the material should be able to sustain plasmonic resonances. Indeed, the calculated optical extinction (see Methods for details) for a PEDOT:OTf nanodisk array (thickness of 30 nm, nanodisk diameter of 500 nm and periodicity of 1000 nm) shows a clear resonance peak at around 2.9 µm (Fig. 1c), which is absent for the non-structured thin film. The small extinction kink at about 1.4 µm disappears when examining single nanodisks instead of arrays (Fig. S4) and hence, arises from lattice scattering of the array (see Fig. S5a). Examining the optical nearfield profile at resonance (2.9 µm) for one of the nanodisks reveals that the extinction peak originates from a dipolar plasmonic mode (Fig. 1d and e), with enhanced fields on the opposite edges of the nanodisk in the polarization direction of the incident light. The optical nearfield patterns slightly above (Fig. 1d) and through the nanodisk (Fig. 1e) both resemble that of traditional gold nanodisk antennas (comparison in Fig. S6). Varying periodicity for fixed nanodisk dimensions did not significantly shift the resonance wavelength (Fig. S5b), further confirming that the extinction peak originates from localized surface plasmons rather than grating effects. In fact, also single nanodisks (Fig. S4a) show the same nanooptical behaviour, with almost identical resonance positions as the periodic arrays (Fig. S4b).

To experimentally verify excitation of plasmons in conductive polymer nanostructures, we fabricated PEDOT:OTf nanodisks on sapphire substrates, using a modified version of colloidal lithography[11] (see Methods and Fig. S7 for details). The protocol could provide large areas of nanodisks of desired diameters, visualized by atomic force microscopy (AFM) for nanodisk diameters of 120 nm, 280 nm, and 710 nm in Fig. 2a, b, and c, respectively (more images are provided in Fig. S8). The nanodisks all originate from 30 nm thick PEDOT:OTf films, while the final thickness of the disks varied somewhat due to residual PMMA remaining on top of the disks after fabrication (Fig. S9). Importantly, the fabricated polymer nanodisk samples exhibit clear extinction peaks (Fig. 2d, e, and f), verifying the simulated nanooptical behaviour. As expected for plasmonic nanoantennas, the resonance positions increase with disk diameter. The experimental results largely match the simulated predictions (Fig. 2g, h, and i) in terms of spectral shapes, peak widths and resonance wavelengths. Small differences in peak positions are attributed to geometrical differences and imperfections of the fabricated nanodisks. The experimental peaks also show somewhat larger broadening, as expected for measured ensembles compared with simulated arrays composed of identical nanostructures[12]. We also note a tendency of slightly more asymmetric peak shape for the fabricated nanodisks compared with the simulated spectra, manifested by a smaller slope on the red side of the peak. As indicated by simulations (Fig. S10), this feature may be related to some few nanometers of remaining PEDOT:OTf film between nanodisks in some samples, although this needs further investigation. The experimental features between 2.7 µm and 3.3 µm (with multiple closely-packed sharp peaks) and at 4.3 µm are due to absorption by water vapour and carbon dioxide[13], respectively, and therefore absent in the simulated spectra.

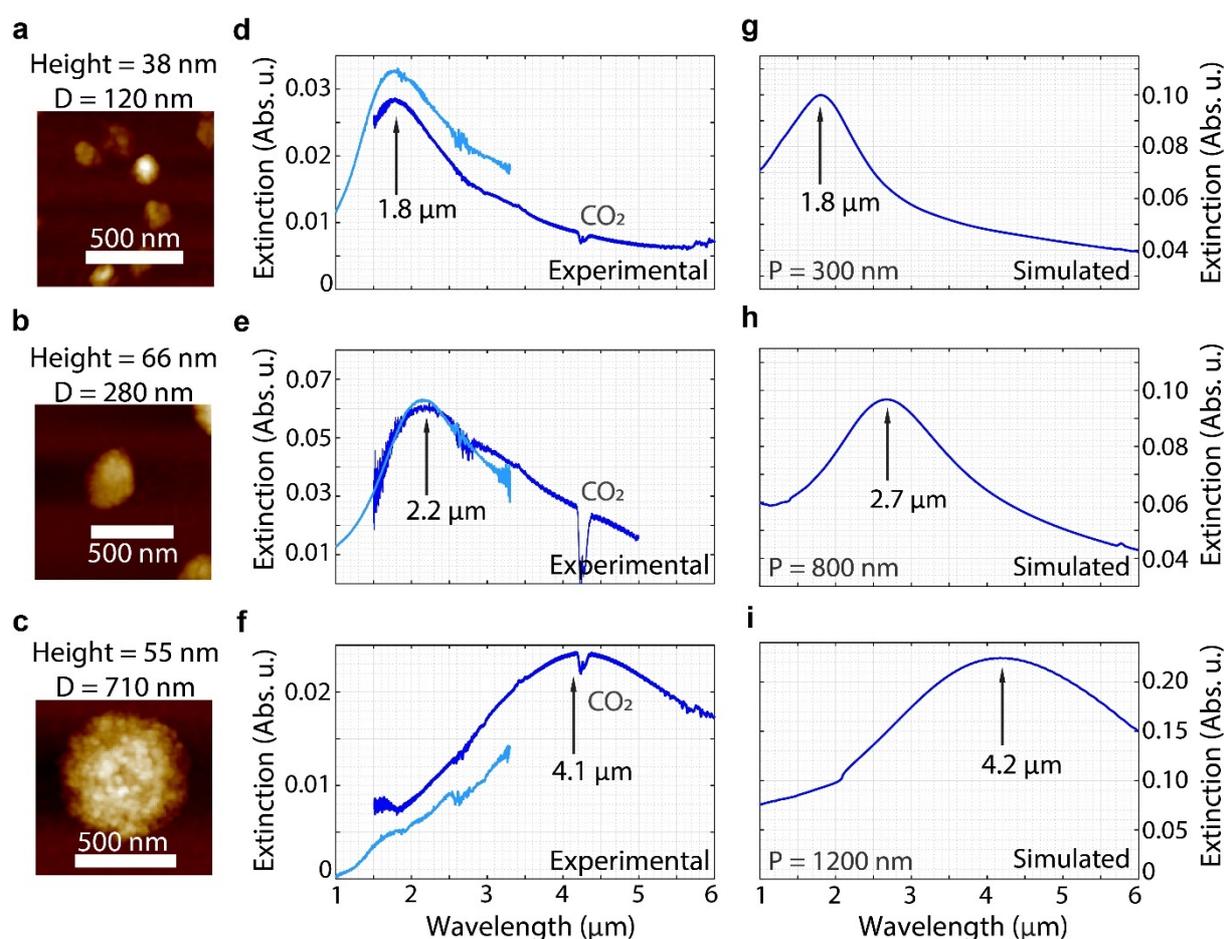

**Fig. 2 | Extinction spectra of PEDOT:OTf nanodisk arrays.** Three different sizes of nanodisk arrays were made on sapphire substrates: **a**, AFM image of a 120 nm diameter nanodisk; **b**, AFM image of a 280 nm diameter nanodisk; **c**, AFM image of a 710 nm diameter nanodisk. The diameter and height measurements are in **Fig. S11**. **d**, **e**, and **f**, Experimental measured extinction spectra of 120 nm, 280 nm, and 710 nm diameter nanodisks. UV-Vis-NIR measurements are plotted in light blue and FTIR measurements are in dark blue. **g**, **h**, and **i**, Simulated extinction spectra of 120 nm, 280 nm, and 710 nm diameter nanodisk arrays. In the simulation, the PEDOT:OTf thickness was 30 nm and the excessive thickness (8 nm, 36 nm, and 25 nm respectively) comes from remaining unremoved PMMA layer.

The results above indicate that the resonance position of the polymer nanodisk antennas can be tuned by geometry, which we here investigate in more detail. Fig. 3a presents the simulated extinction for 30 nm thick single PEDOT:OTf nanodisks of varying diameter on a substrate with refractive index of 1.6. Normalized extinction versus diameter is presented in Fig. 3c as colour maps. It is clear that the resonance position redshifts with increasing diameter, enabling tuning in a large spectral range from around 2 µm to around 4 µm for disks with sizes ranging from 200 nm and 700 nm in diameter. The spectral tunability can likely be extended further by other geometries. While the nanodisk resonances redshift with increasing disk diameter, they instead blueshift with increasing thickness, as presented in Fig. 3d and 3f for nanodisks with fixed diameter of 500 nm. Both these geometrical dependencies match expectations based on plasmonic nanodisk resonances[14-16].

To enable analytical calculation of the optical response, we approximate the nanodisks as oblate spheroids with diameter $D$ and thickness $t$, which in the quasi-static limit $D \ll \lambda$ gives the dipolar polarizability $\alpha$ as[17]

$$\alpha(\lambda) = V \frac{\varepsilon(\lambda) - \varepsilon_s}{\varepsilon_s + L[\varepsilon(\lambda) - \varepsilon_s]} \qquad (1)$$

where $V$ is the volume of the spheroid and $\varepsilon_s$ is the permittivity of the surrounding medium. We use the in-plane permittivity of PEDOT:OTf as $\varepsilon(\lambda)$ and set $\varepsilon_s$ = 1.69 as the effective surrounding permittivity for disks in air on a substrate with refractive index 1.6 (see Methods). $L$ is a geometrical factor that equals 1/3 for a sphere ($D = t$) and decreases for increasing nanodisk ratio ($D > t$, see Fig. S12). To fulfil the resonance condition of maximum polarizability when $L$ decreases, the magnitude of the negative permittivity needs to increase. Because the permittivity of the conductive polymer increases in magnitude with wavelength (see Fig. 1b), the resonance position therefore redshifts with increasing aspect ratio ($D/t$). This illustrates why the resonance of the PEDOT:OTf nanoantennas redshifts with increasing disk diameter and blueshifts with increasing disk thickness. Larger disks require corrections for finite wavelength effects, which gives the corrected polarizability as[18,19]

$$\alpha'(\lambda) = \alpha(\lambda)\left[1 - \frac{k^2}{2\pi D}\alpha(\lambda) - i\frac{k^3}{6\pi}\alpha(\lambda)\right] \qquad (2)$$

where $k$ is the wave number of the incident light. The extinction cross-section $\sigma(\lambda)$ can now be calculated via[17]

$$\sigma(\lambda) = k\text{Im}[\alpha'(\lambda)] \qquad (3)$$

Fig. 3b and 3e show the final calculated extinction cross sections of single PEDOT:OTf oblate spheroids, with sizes corresponding to the simulated nanodisks in Fig. 3a and 3d, respectively. The calculated extinction based on the dipolar polarizability match the results from the full simulations quite well, both in terms of extinction magnitude and its increase with aspect ratio, and in terms of peak positions and their redshift with increasing nanodisk aspect ratio. The results thereby further corroborate that the observed extinction peaks of the PEDOT:OTf nanodisks originate from dipolar plasmonic excitations.

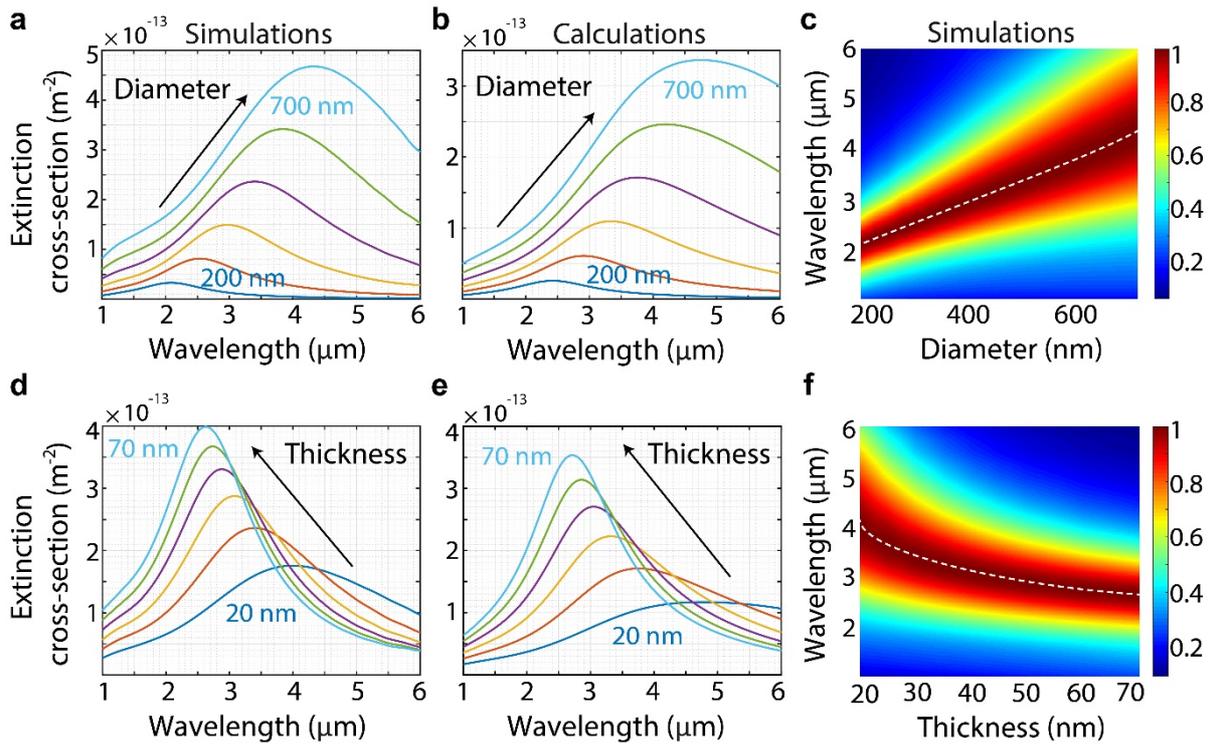

**Fig. 3 | Geometry dependence of single PEDOT:OTf nanodisk localized plasmons. a**, Simulated extinction cross-section of 30 nm thick single nanodisks of different diameters (diameter step size = 100 nm, substrate refractive index = 1.6). **b**, Analytically calculated extinction cross-section for oblate spheroids corresponding to the nanodisk sizes in **a**. **c**, Colour map of simulated normalized extinction versus diameter for single 30 nm thick nanodisks. **d**, Simulated extinction cross-section of 500 nm in diameter single nanodisks with different thickness (thickness step size = 10 nm, substrate refractive index = 1.6). **e**, Analytically calculated extinction cross section for oblate spheroids corresponding to the nanodisk sizes in **d**. **c**, Colour map of simulated normalized extinction versus diameter for single 30 nm thick nanodisks. **f**, Colour map of simulated normalized extinction versus thickness for single 500 nm in diameter nanodisks. The white dashed lines in **c** and **f** indicate the resonance peak position and its shift with changes in diameter and thickness, respectively. The color scale bars in **c** and **f** present the normalized extinction. See **Fig. S13** for normalized extinction spectra corresponding to **a** and **d**.

Finally, we demonstrate that the conductive polymer nanoantennas can be switched on and off. Such dynamic control of optical nanoantennas is a topic of intense research, not least motivated by the enormous potential for important applications enabled by active plasmonic devices and metasurfaces[4]. Among various approaches to tune nanophotonic systems, recent research has explored tuning by modulating the free charge carriers in plasmonic systems, including electrical gating[20] and photo-carrier excitation[21]. While this approach is rather limited for traditional plasmonic materials, conductive polymers hold great promise since their bipolaron charge carrier concentration can be modulated by several orders of magnitude via their redox state[5]. Here, we control the redox state chemically, utilizing that exposure of PEDOT:OTf to highly branched PEI (poly[ethylene imine], see chemical structure in left panel of Fig. 4a) vapour (volatile impurities of ethyleneimine dimers and trimers) can reduce the material[22]. Optical extinction spectroscopy of a (non-structured) thin PEDOT:OTf films visualizes the process via almost complete reduction of the free charge carrier absorption in the IR and the emergence of a neutral state peak at around 600 nm (see Fig. 4b)[22]. For the reduced polymer, the reaction between PEI impurities and counterions reduces the bipolaronic charge carrier concentration, resulting in a material with largely reduced electrical conductivity (schematic mechanism in Fig. 4a right panel)[22]. The process is reversible and we can recover the original optical properties of the PEDOT film via acid treatment of the reduced film (see Methods). This process re-oxidizes the material, for which the neutral state disappears and the absorption returns to that of the initial pristine film (Fig. 4b). Knowing that the optical material properties of PEDOT:OTf can be reversibly modulated, we utilize this feature to actively tune our polymer nanodisk metasurfaces. The black curve in Fig. 4c shows the extinction spectra of a PEDOT:OTf nanodisk array in its oxidized pristine state, with plasmonic resonance peak at around 1900 nm. This peak completely disappears upon PEI vapour treatment, for which the material in the nanodisks is no longer plasmonic, due to drastic reduction of the bipolaronic charge carrier concentration. Indeed, the neutral state material absorption emerges at 600 nm for the PEI vapour treated metasurfaces. Importantly, the optical properties are not volatile, but stable over time and we observe only minimal extinction changes of the sample after one week (see Fig. S14). By re-oxidizing the sample with sulfuric acid, the plasmonic resonance peak recovers to its initial state, with both similar intensity and width as for the original plasmonic metasurface (Fig. 4c). The increase in extinction below 800 nm for the re-oxidized sample is likely due to different probe areas combined with some polystyrene beads remaining after fabrication (similar effects were observed for samples before and after bead removal, Fig. S15).

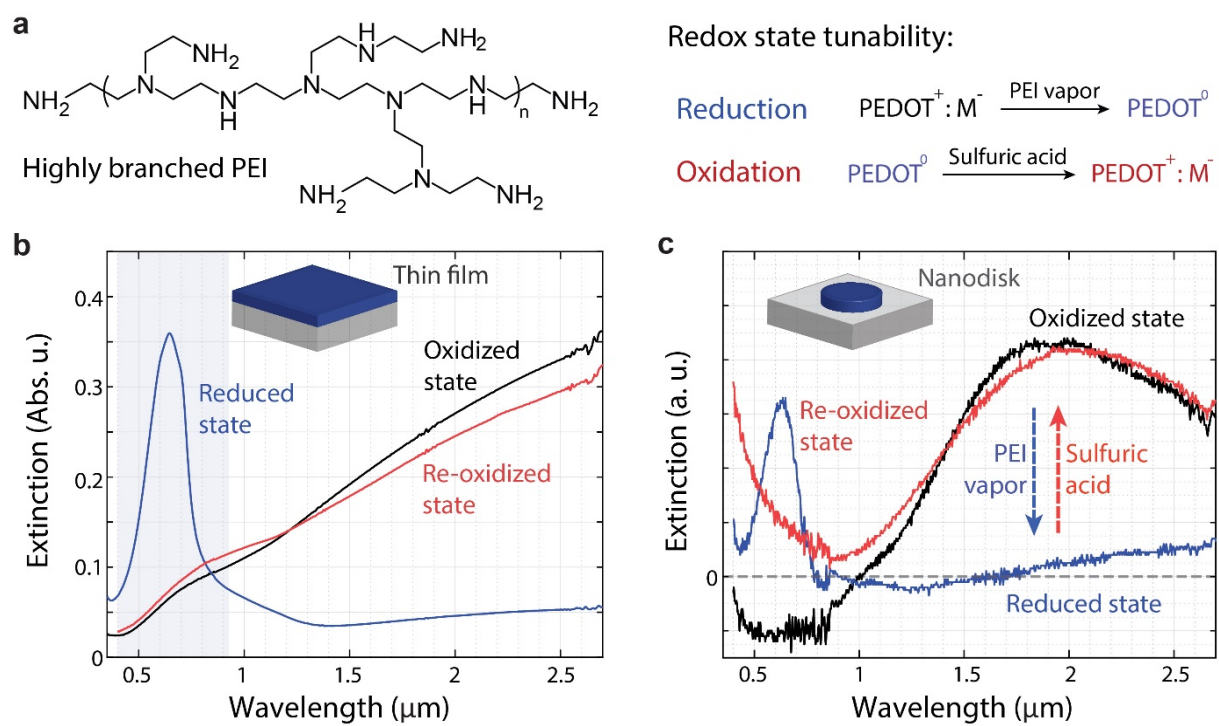

**Fig. 4 | Redox state tunability of PEDOT:OTf nanodisk arrays. a**, Chemical structure of highly branched PEI (left panel) and the redox state tunability mechanism for PEDOT-based materials (right panel) where M$^-$ stands for the counterion of PEDOT (e.g. PSS, Tos, or OTf). **b**, Measured extinction for thin PEDOT:OTf film on glass at different redox states (black: oxidized state, blue: reduced state, and red: re-oxidized state). Shaded area indicates the neutral state absorption peak. **c**, Measured extinction for PEDOT:OTf nanodisk arrays on glass (thickness of 43 nm and nanodisk diameter of 140 nm) at different redox states (black: oxidized state, blue: reduced state, and red: re-oxidized state).

We have demonstrated that nanodisks made of highly conductive polymers can be used as optical nanoantennas to form active plasmonic metasurfaces. While previous research has showed excitonic resonances in organic nanomaterials made from small molecules and dyes,[23,24] our system is fundamentally different in that the resonances originate from localized plasmons formed by bipolaronic charge oscillations. Our observations open up new avenues for dynamically controllable plasmonics based on redox-tunable conductive polymers, where future work may explore dynamic control also by other means, including electrochemical modulation of the PEDOT:OTf redox state[25]. The future also holds promise for conductive polymers with yet further improved plasmonic properties, based on strategies to improve electrical conductivity and lower defect density, including effective chain alignment[26], and sequential doping[27]. We hope that our study of redox-tunable conductive polymer plasmonics will inspire research in this interdisciplinary field of manipulating light-matter interaction with organic materials at the nanoscale.

## Acknowledgements

The authors thankfully acknowledge financial support from the Swedish Research Council, the Swedish Foundation for Strategic Research, the Wenner-Gren Foundations, and the Swedish Government Strategic Research Area in Materials Science on Functional Materials at Linköping University (Faculty Grant SFO-Mat-LiU No. 2009 00971).

## Author contributions

M.P.J. conceived and supervised the project. S.C, V.S., P.K., and V.D. performed ellipsometry measurements and data analysis. S.C. and M.S.C. fabricated the nanostructures. S.C., M.P.J. and E.S.K. performed numerical simulations. S.C. and H.S. performed PEI vapour treatments. S.C performed all the other characterizations. S.C. and M.P.J. organized the data and wrote the manuscript. All authors reviewed and commented on the manuscript.

## Competing interests

The authors declare no conflicts of interest.

## Methods

**Thin film deposition.** PEDOT:OTf thin film were prepared *via* vapour phase polymerization (VPP) as reported in the literature.[8,28] The oxidant solution for EDOT polymerization was prepared by mixing 0.03 g of iron (III) trifluoromethanesulfonate (Fe[OTf]3, from Alfa Aesar), 0.2 g of tri-block co-polymer poly(ethylene glycol)-block-poly(propylene glycol)-block-poly(ethylene glycol) (PEG-PPG-PEG or P-123, average $M_n$ ~5,800, from Sigma-Aldrich) and 0.8 g of 99.5% ethanol (from Solveco). Oxidant films were deposited by spin-coating at 1500 rpm for 30 s onto pre-cleaned sapphire or glass substrates (sonicated in cleaning detergent, de-ionized water, acetone, and isopropanol each for 10 min respectively and treated with oxygen-plasma at 200 W for 5 min before use). After 30 s baking on a hotplate at 70 °C, the samples were transferred into a heated vacuum desiccator (Vacuo-temp, from SELECTA). EDOT (142.18 g mol$^{-1}$, from Sigma-Aldrich) droplets were drop-casted onto a glass substrate on a hot plate at 30 °C in the desiccator to ensure its evaporation. After 30 min of polymerization at a pressure of 70 mBar, the samples were taken out from the desiccator and washed with ethanol multiple times to remove byproducts and unreacted residues, followed by air-drying with nitrogen. Acid treatment by soaking in 1 M sulfuric acid ($H_2SO_4$) for 10 min at 100 °C was applied to the samples for further enhancement of their electronic properties.[29] With acid treatment, parts of counterions (OTf) in PEDOT:OTf thin film are replaced by $SO_4^{2-}$ from the sulfuric acid.

**Nanoantenna fabrication.** The detailed process flow for nanodisk array fabrication is shown in **Fig. S7**, which is a modified version of colloidal lithography.[11] Briefly, a 4 wt% PMMA (Mw ~996,000, from Sigma-Aldrich) solution in anisole (from Sigma Aldrich) was spin-coated onto the as-prepared PEDOT:OTf thin films. Soft baking at 140 °C for 10 min was then applied. The samples were treated with reactive oxygen plasma (50 W, 250 mTorr) for 5 s to increase the hydrophilicity of the surface. In order to functionalize the PMMA surface to be positively charged, 2 wt% poly(diallyldimethylammonium chloride) (PDDA, 522376 from Sigma-Aldrich) in DI water was dropped on the samples. After 1 min, the samples were rinsed with deionized water for 40 s and dried with nitrogen stream. Negatively charged polystyrene nanoparticles (PS beads with different diameters, 0.2-0.3 wt% in deionized water, from Microparticles GmbH) were then dropped on the samples. After 10-30 min, the samples coated with PS beads were rinsed with DI water and dried with nitrogen stream resulting in a sparse monolayer of PS beads on the PMMA/PEDOT:OTf thin films. A heat treatment at 100 °C for 2 min were applied to the samples to improve the adhesion of PS beads on the samples. Reactive oxygen plasma etching (250 mTorr, 50 W) for 3-5 min were applied to the samples, using the PS beads monolayer as mask. Depending on the size of PS beads and thickness of PMMA and PEDOT:OTf thin films, the time interval of etching can be varied to ensure a complete removal of PMMA and PEDOT:OTf parts that are not covered by the mask. The samples were then placed into an acetone bath and soaked for 10-30 min followed by a mild sonication for 3 min and nitrogen stream drying to remove PMMA and PS beads and finally the PEDOT:OTf nanodisks were obtained. In this study, three different diameters of PS beads were used: 239 nm (PS-ST-0.25, Microparticles GmbH), 497 nm (PS-ST-0.50, Microparticles GmbH), and 1046 nm (PS-ST-1.0, Microparticles GmbH).

**Vapour treatment of thin films and nanoantennas.** The vapour treatment was conducted inside a $N_2$-filled glovebox by exposing the samples to the vapour of ethyleneimine dimers and trimers by heating a vial containing highly branched poly(ethylene imine) liquid (PEI, $M_w$ ~ 800, from Sigma-Aldrich) at 120 °C for 5 min.[22] After the vapour treatment, the samples were annealed at 120 °C for another 5 min. For re-oxidizing of the samples, they were put into 1 M sulfuric acid bath for 10 min followed by a drying process of 10 min at 100 °C on a hot plate.

**Ellipsometry.** PEDOT:OTf thin film samples were measured at normal ambient conditions at room temperature. The films were deposited on 2-inch single side polished *c*-plane sapphire wafers (from Semiconductor Wafer Inc.). Ellipsometric data for PEDOT:OTf thin films were collected using three different ellipsometers covering a wide spectral range from 0.0028 eV to 5.9 eV. UV-Vis-NIR measurements were performed on a J. A. Woollam Co. RC2® spectroscopic ellipsometer for five incident angles (40°, 50°, 60°, 70°, and 80°) and spectral range from 0.73 eV (1690 nm) to 5.90 eV (210 nm). Infrared measurements were performed on a J. A. Woollam Co. IR-VASE® spectroscopic ellipsometer for two incident angles (50° and 70°) and spectral range from 28.0 meV (230 cm$^{-1}$) to 1.0 eV (7813 cm$^{-1}$). THz measurements were performed on the THz ellipsometer at the Terahertz Materials Analysis Center (THeMAC) at Linköping University.[30] Three incident angles (40°, 50°, and 60°) were used for THz measurements, in the spectral range between 2.8 meV (0.67 THz) and 4.0 meV (0.97 THz). The typical ellipsometer measures the complex reflectance ratio $\rho$ at different frequencies, as obtained from $\rho = r_p/r_s = tan(\Psi)e^{i\Delta}$, where $r_p$ and $r_s$ are the complex Fresnel reflection coefficients for p- and s-polarized light; $\Psi$ shows the amplitude ratio change of the two polarizations; and $\Delta$ indicates the phase difference between them.[31] WVASE® (J. A. Woollam Co.) was used for data analysis and a Drude-Lorentz model were employed for model fitting and optical parameter extraction for the PEDOT:OTf thin films.[8]

**Electrical and structural characterization.** Sheet resistance, $R_s$, of the thin film was measured using a 4-point probe set-up using a Signatone Pro4 S-302 resistivity stand and a Keithley 2400. Film thickness *t* was determined by a surface profiler (Dektak 3st, Veeco). The thickness of the VPP PEDOT:OTf films varied in the range from 30 to 40 nm. The electrical conductivity can then be calculated by $\sigma = 1/(R_s t)$. Atomic force microscopy (AFM) was employed for surface morphology characterization, in tapping mode using a Veeco Dimension 3100. The morphological images were analyzed using Nanoscope Analysis software (Bruker).

**Optical characterization.** The extinction spectra in the Vis-NIR range (400 nm to 3300 nm) were measured using a UV-Vis-NIR spectrometer (Lambda 900, Perkin Elmer Instruments). The extinction spectra include transmission losses due to both absorption and scattering. Fourier-transform infrared spectroscopy (FTIR) measurements were performed in the spectral range from 1333 nm (7500 cm$^{-1}$) to 5000 nm (2000 cm$^{-1}$) or 6667 nm (1500 cm$^{-1}$) using an Equinox 55 spectrometer (Bruker). FTIR spectra were acquired in absorbance mode using a resolution 4 cm$^{-1}$ and 100 scans. Samples deposited on 20×20×0.5 mm double-side polished sapphire substrates (from Semiconductor Wafer Inc.) were made for FTIR and UV-Vis-NIR measurements.

**Optical numerical simulations.** Numerical simulations (electric nearfield intensity and farfield spectra) of the electromagnetic response of PEDOT:OTf nanoantennas were performed *via* the finite-difference time-domain (FDTD) method using the commercial software Lumerical FDTD Solutions (http://www.lumerical.com/fdtd.php). The optical parameters for the PEDOT:OTf thin film were taken as the anisotropic complex permittivity obtained from the ellipsometry measurements. For periodic nanodisk arrays and thin films, the spectra and nearfield profiles were recorded via field and power monitors. Periodic PEDOT:OTf nanodisk arrays (or thin film) were placed on top of glass or sapphire substrates. The structures were illuminated by a planewave light source at normal incidence. Anti-symmetrical and symmetric boundaries were used for the *x*-axis (parallel to polarization) and *y*-axis (normal to polarization) and perfectly matched layer (PML) were used for the *z*-axis (parallel to light incident direction). For single nanodisks, spectra were obtained using a total field/scattered field and by extracting the extinction cross-section of isolated PEDOT:OTf nanodisks on a sapphire substrate. Geometry parameters are indicated in each graph (diameter, thickness, and periodicity) and the mesh size was typically 3 × 3 × 3 nm$^3$, or 2 × 2 × 2 nm$^3$ for the smaller size disks. The optical parameters for gold[32], glass[33] and PMMA[34,35] were taken from literature while the permittivities of sapphire substrate and PEDOT:OTf were determined by ellipsometry. In the analytical calculations, the effective permittivity of the surroundings was calculated based on an average refractive index of air and sapphire ($\varepsilon_s = [(n_{air} + n_{sapphire})/2]^2$). The refractive index of sapphire is 1.75 at about 1 μm and 1.6 at about 5 μm and for simplicity we fix $n_{sapphire}$=1.6 which gives $\varepsilon_s$=1.69.

# Supporting Information

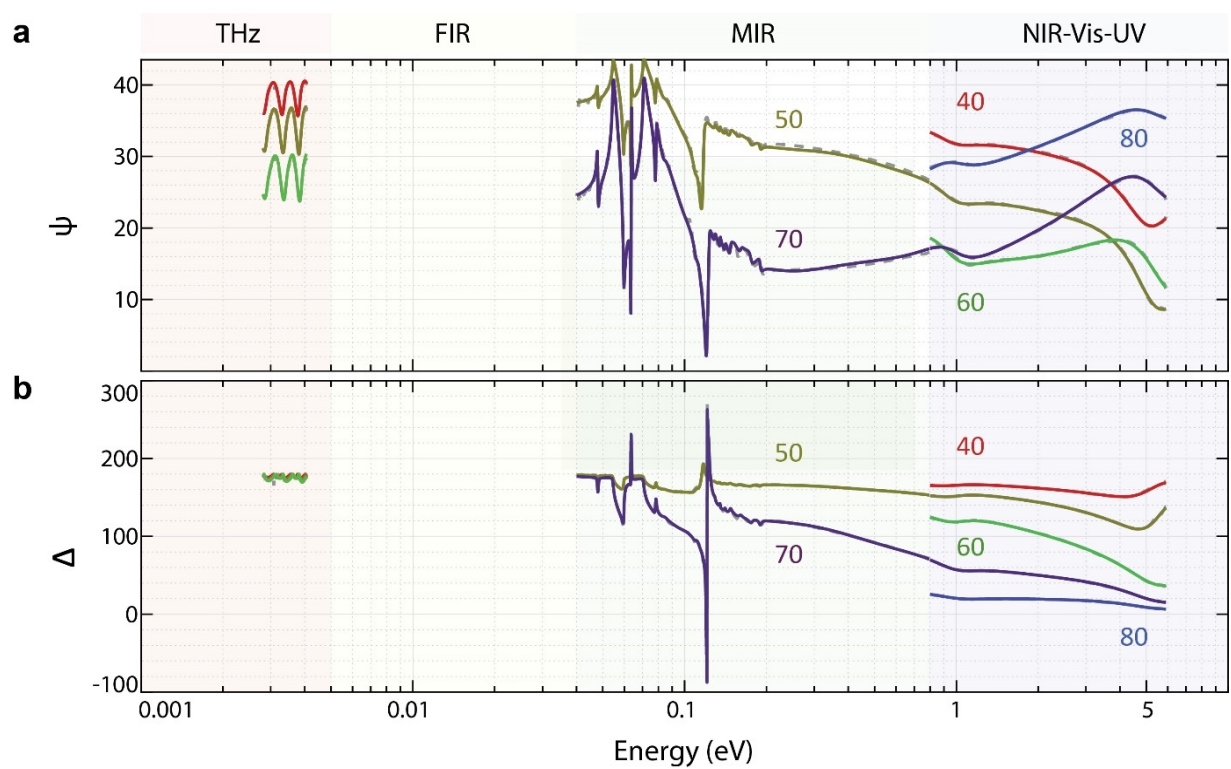

**Fig. S1 | Spectroscopic ellipsometry for thin PEDOT:OTf film.** Ellipsometric raw data **a**, $\psi$ and **b**, $\Delta$ of the film are exhibited with the spectral range from 2.8 meV (0.67 THz) to 5.90 eV (210 nm), including THz, MIR, and NIR-Vis-UV three regions. For each spectral window, two, three or five incident angles were employed. The experimental data are in grey dashed lines and Drude-Lorentz model[8] generated data are in solid lines. Details of the model can be found in **Table. S1**.

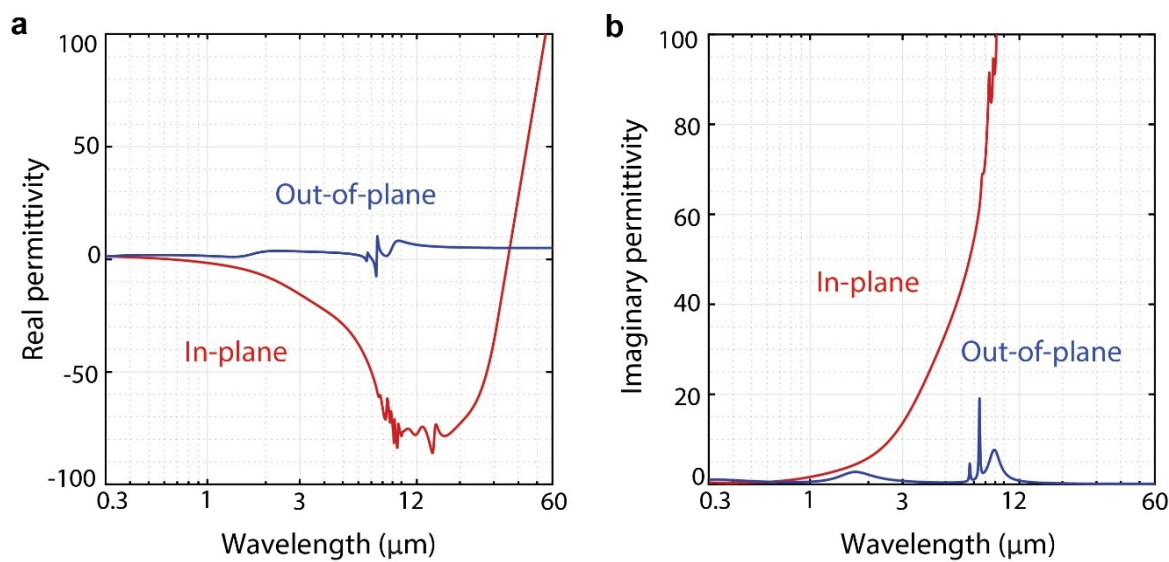

**Fig. S2 | Permittivity dispersion for thin PEDOT:OTf thin film.** Both **a**, real permittivity, and **b**, imaginary permittivity dispersion curves are plotted. Red curves are for the in-plane direction (*x-y* plane) and blue curves for the out-of-plane direction (*z*-axis). The extracted data are based on the Drude-Lorentz model in **Fig. S1** and using the fitted parameters in **Table. S1**.

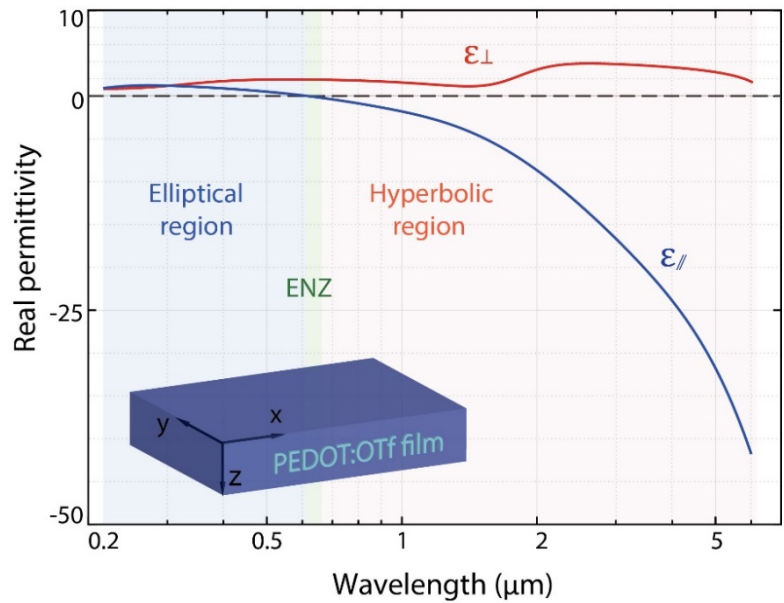

**Fig. S3 | Natural hyperbolic properties of a PEDOT:OTf thin film.** In-plane ($\varepsilon_\perp$, ordinary axis or *x-y* plane) and out-of-plane ($\varepsilon_\parallel$, extraordinary or *z*-axis) real permittivity comparison. Three spectral regions are indicated: elliptical regime (permittivity for both directions are positive), hyperbolic regime (the sign of permittivity for the two directions are opposite), and ENZ regime (epsilon-near-zero, one direction has permittivity close to zero). The results show that the PEDOT:OTf thin film has a natural hyperbolic permittivity in the range 0.7 µm and 6 µm.

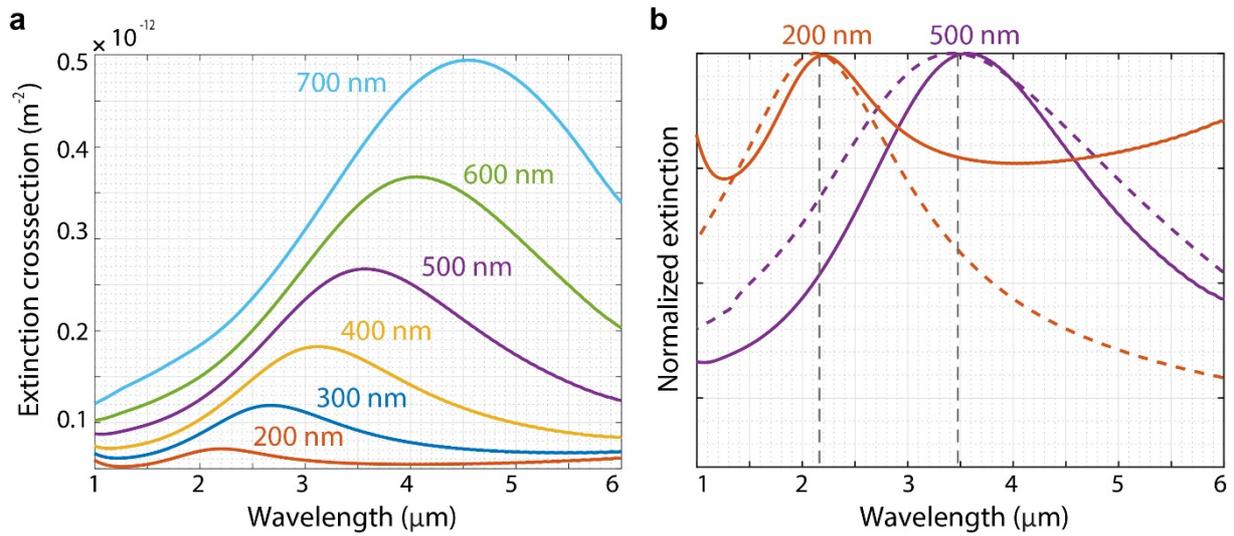

**Fig. S4 | Simulated optical extinction spectra for single nanodisks. a**, The extinction spectra for single 30 nm thick nanodisks with different diameters from 200 nm to 700 nm on sapphire substrates. **b**, Comparison between normalized single nanodisk extinction (solid curves) and normalized periodic nanodisk arrays extinction (dashed curves). The resonance peak positions are similar for single and periodic nanodisks.

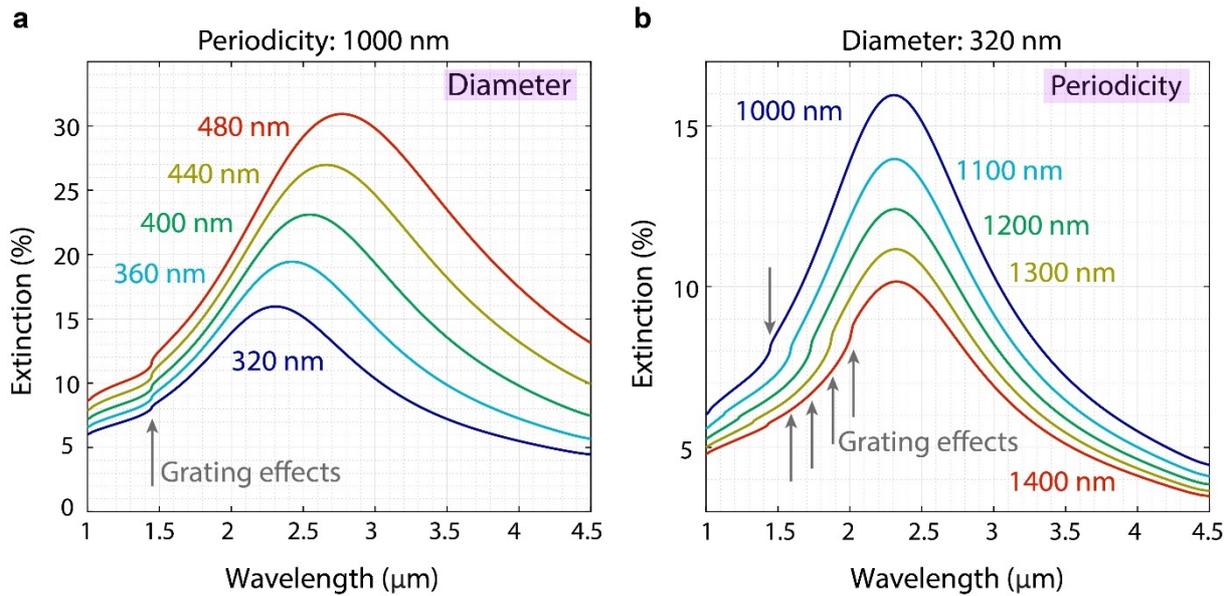

**Fig. S5 | Grating effects in periodic PEDOT:OTf nanodisk arrays.** PEDOT:OTf nanodisk arrays with thickness of 40 nm on glass substrates, with 100 nm PMMA on top of the nanodisks. **a**, Extinction of nanodisks with different diameters (from 320 nm to 480 nm) at a fixed periodicity of 1000 nm. The kink feature at about 1.4 μm comes from grating effects. **b**, Extinction of nanodisks with different periodicity (from 1000 nm to 1400 nm) at a fixed disk diameter of 320 nm.

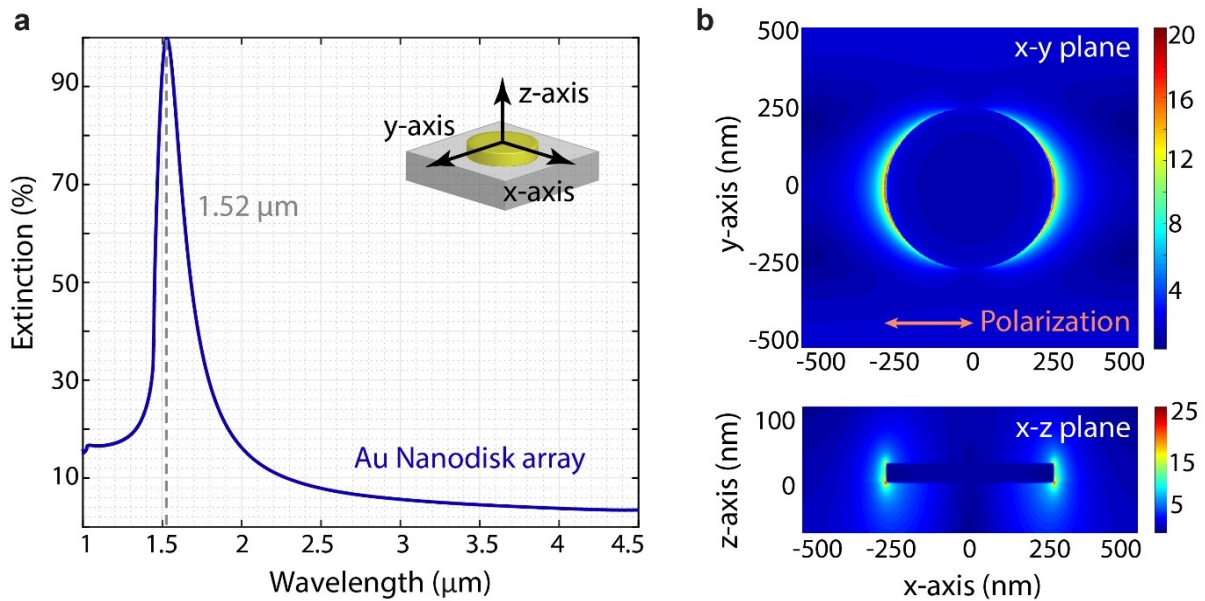

**Fig. S6 | Optical extinction spectra and nearfield profile for gold (Au) nanodisk antennas. a**, Simulated extinction spectra of a Au nanodisk array (thickness of 30 nm, disk diameter of 500 nm, and periodicity of 1000 nm) on a glass substrate. A sharp resonance peak located at 1.52 μm can be observed. *x*-, *y*-, and *z*-axes are indicated in the inset. Calculated nearfield profiles at the wavelength of extinction maximum (1.52 μm) for one of the Au nanodisks of the array in **a**: **b**, *x-y* in-plane direction 2 nm above the nanodisk; **c**, *x-z* out-of-plane direction, cross-section through the center of the nanodisk. The color scale bars show the electric field strength relative to the incident light.

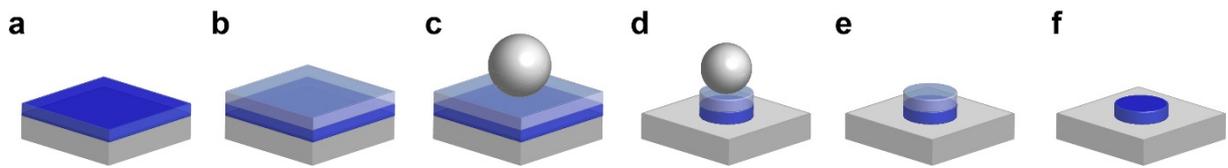

**Fig. S7 | Fabrication procedure of large scale PEDOT:OTf nanodisk arrays.** The process flow is based on a modified version of colloidal lithography[11]. **a**, Deposition of thin PEDOT:OTf film by VPP. **b**, Spin coating of sacrificial PMMA layer on top of PEDOT:OTf film. **c**, Deposition of colloidal PS microbeads on oxygen-plasma treated PMMA layer by colloidal lithography. **d**, Reactive ion etching of PMMA and PEDOT:OTf thin film. The PS microbeads also shrink somewhat. **e**, Tape stripping of PS microbeads. **f**, PMMA layer removal by acetone soaking and sonication.

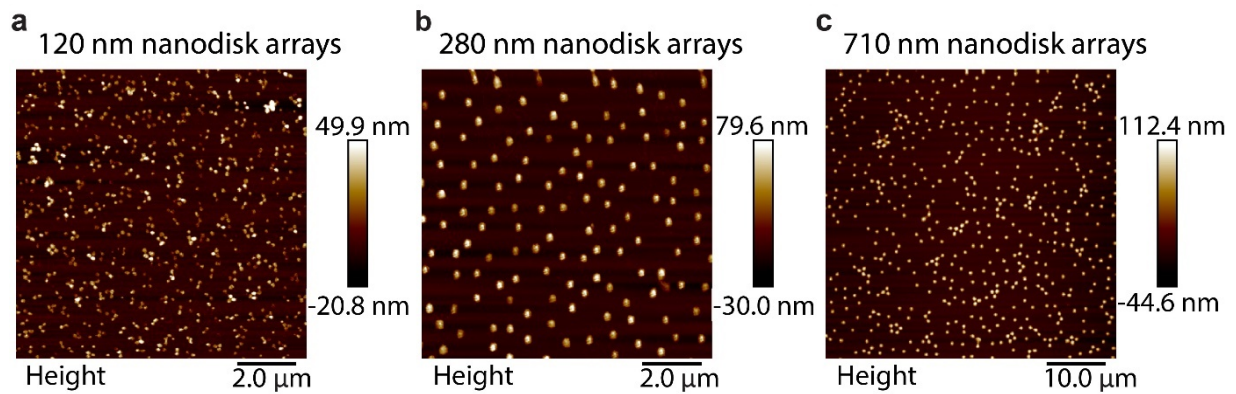

**Fig. S8 | AFM topography images of PEDOT:OTf nanodisk array for three different disk sizes.** **a**, Sample with 120 nm diameter nanodisks, corresponding to **Fig. 2a**. **b**, Sample with 280 nm diameter nanodisks, corresponding to **Fig. 2b**. **c**, Sample with 710 nm diameter nanodisks, corresponding to **Fig. 2c**.

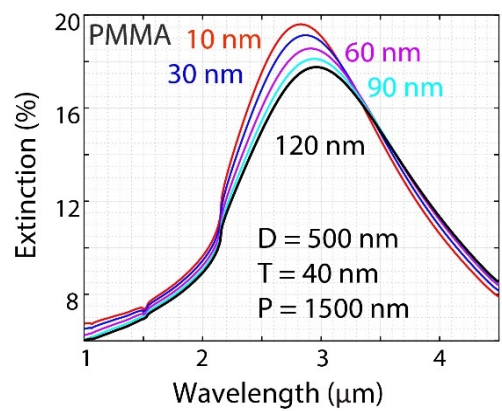

**Fig. S9 | Effects from PMMA layer on PEDOT:OTf nanodisks.** In the simulation, PMMA nanodisk layers with different thickness (10 nm to 120 nm) were added to study its influence on the resonance position. The PEDOT:OTf nanodisk arrays have a thickness (T) of 40 nm, diameter (D) of 500 nm, and periodicity (P) of 1500 nm. With the increase of the PMMA layer thickness, the resonance peak red shifts with a reduced maximal extinction.

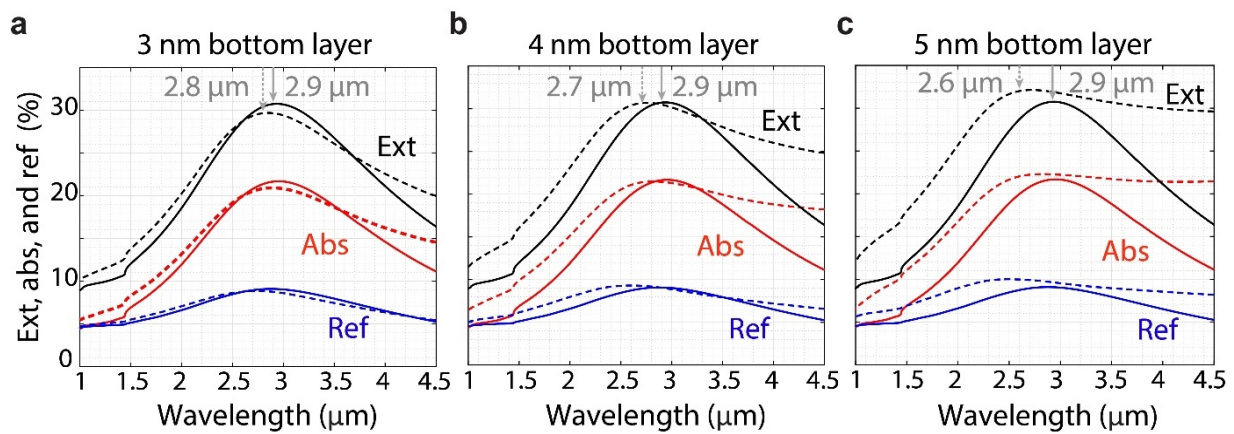

**Fig. S10 | Influence of ultrathin PEDOT:OTf bottom layer for nanodisk arrays. a**, Simulated extinction of 3 nm thick bottom PEDOT:OTf layer on glass substrate. **b**, Simulated extinction of 4 nm thick bottom PEDOT:OTf layer on glass substrate. **c**, Simulated extinction of 5 nm thick bottom PEDOT:OTf layer on glass substrate. For the simulated nanodisk arrays, the thickness is 30 nm, the diameter is 500 nm, and the periodicity is 1000 nm. Samples with bottom layer are plotted in dashed curves and samples without bottom layer are in solid curves. Extinction (Ext, in black), absorption (Abs, in red), and reflection (Ref, in blue) are indicated. For samples with bottom layer thickness less than 3 nm, the resonance peak blue-shift is negligible.

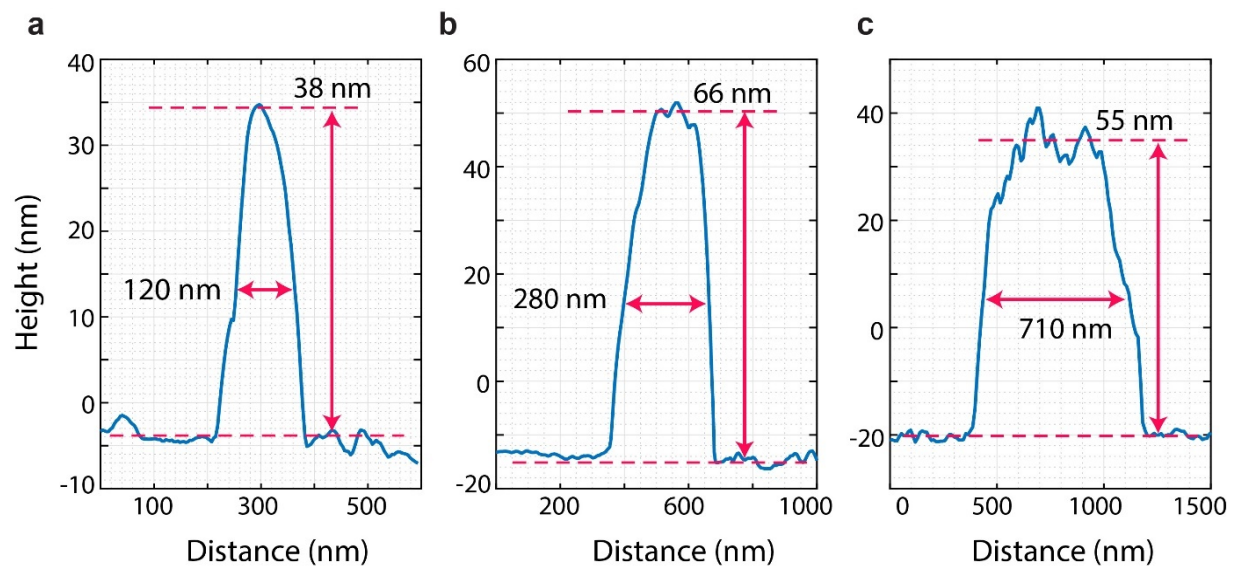

**Fig. S11 | Height profile of single nanodisks.** Height and diameter of PEDOT:OTf single nanodisks measured by AFM. **a**, From Fig. **2a** (height: 38 nm, diameter: 120 nm); **b**, From **Fig. 2b** (height: 66 nm, diameter: 280 nm); and **c**, From **Fig. 2c** (height: 55nm, diameter: 710 nm).

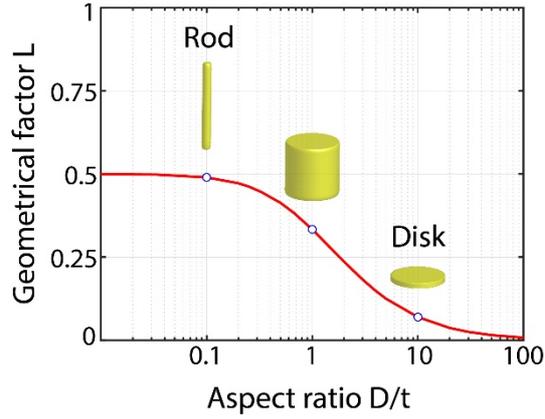

**Fig. S12 | Geometrical factor *L* for different aspect ratios of nanodisks when approximated as oblate spheroids.** The geometrical factor *L*, for excitation at normal incidence, was calculated via[17]

$$L = \frac{D^2 t}{16} \int_0^\infty \frac{dq}{\left(\frac{D^2}{4} + q\right) f(q)} \quad (E1)$$

$$f(q) = \sqrt{\left(q + \frac{D^2}{4}\right)^2 \left(q + \frac{t^2}{4}\right)} \quad (E2)$$

where *D* and *t* are the diameter and thickness of the nanodisks. With the increase of aspect ratio (ratio between diameter and thickness of the disk, *D/t*), the geometrical factor *L* decreases from 0.5 for long rods to 1/3 for *D* = *t*, and further towards 0 for ultrathin disks.

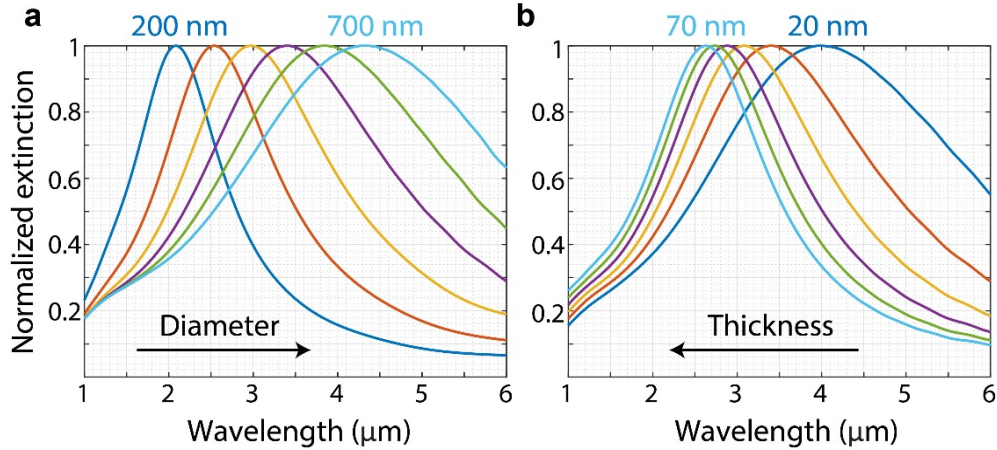

**Fig. S13 | Normalized extinction cross-section spectra for single PEDOT:OTf nanodisks of different dimensions. a**, Spectra for varying diameter show increase in resonance wavelength with increasing diameter (from 200 to 700 nm, with thickness fixed at 30 nm). **b**, Spectra for varying disk thickness show decrease of resonance wavelength with increasing thickness (from 20 to 70 nm, with diameter fixed at 500 nm. The spectra were simulated based on single PEDOT:OTf nanodisks on a substrate with fixed refractive index of 1.6.

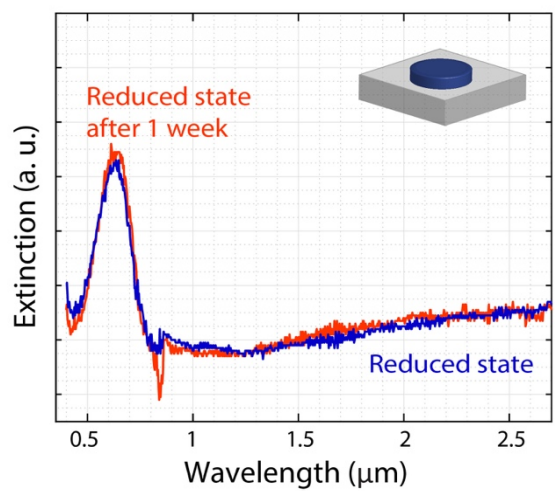

**Fig. S14 | Extinction of reduced PEDOT:OTf nanodisks.** The optical properties of as-prepared nanodisks (blue curve) is stable and changes little after one week in air (red curve).

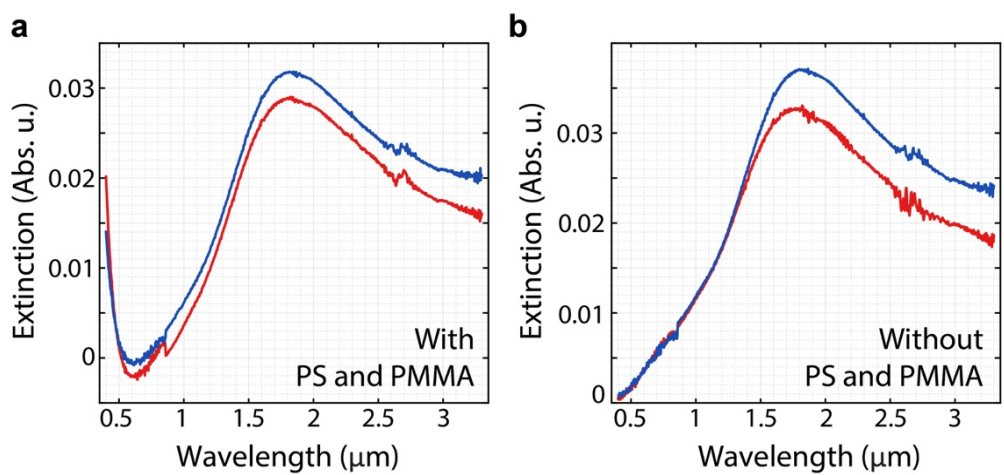

**Fig. S15 | Measured extinction spectra for PEDOT:OTf nanodisks with and without PMMA and polystyrene beads on top of the nanodisks.** The two different colors correspond to two different samples, both with nanodisk diameter of 120 nm. **a**, Extinction spectra for nanodisks with polystyrene beads remaining on top of the disks. A strong absorption can be observed below 700 nm due to the polystyrene beads. **b**, Extinction spectra for PEDOT:OTf nanodisks after removing the polystyrene beads, for which the absorption feature of the polystyrene below 700 nm has disappeared.

**Table. S1 | Measured electrical properties of thin PEDOT:OTf film.** The parameters were determined by both ellipsometry and conventional electrical measurements (profiler and 4-point-probe). The predictions from ellipsometry are based on the anisotropic Drude-Lorentz model[8].

|  | Ellipsometry predictions | Experimental measurements |
|---|---|---|
| **Thickness (nm)** | 31.802 ± 0.394 | 30 ± 5 |
| **Electrical conductivity (S/cm)** | 5603.8 ± 760.0 | 5200 ± 500 |
| **Charge density ($10^{21}$ cm$^{-3}$)** | 2.61 ± 0.07 |  |
| **In-plane mobility (cm$^2$ V$^{-1}$ s$^{-1}$)** | 13.419 ± 1.46 |  |